\documentclass[10pt]{article}
\usepackage{latexsym,graphicx}
\newcommand{\be}{\begin{equation}}
\newcommand{\ee}{\end{equation}}
\def\n{\noindent}
\catcode `\@=11
\catcode `\@=12
\begin{document}
\begin{center}
\large{\bf {Bulk Viscous cosmological models in Lyra geometry}} \\
\vspace{10mm}
\normalsize{\bf {Anirudh pradhan $^1$ and Hare Ram Pandey$^2$}} \\
\normalsize{$^{1}$Department of Mathematics, Hindu Post-graduate College,\\
 Zamania-232 331, Ghazipur, India} \\
\normalsize{E-mail: pradhan@iucaa.ernet.in, acpradhan@yahoo.com}\\
\normalsize{$^2$Department of Mathematics, R. S. K. I. College,\\
Dubahar, Ballia- 277 405, India}
%\normalsize{}
\end{center}
\vspace{10mm}
%\date{}
%\maketitle
\begin{abstract} 
{We have investigated an LRS Bianchi Type I models with bulk viscosity
in the cosmological theory based on Lyra's geometry. A new class of 
exact solutions have been obtained by considering a time-dependent 
displacement field for a constant value of the deceleration parameter
and viscosity coefficient of bulk viscous fluid is assumed to be a power 
function of mass density. The physical behaviour of the models is also 
discussed.}{}{} 
\end{abstract}
\smallskip
%\begin{flushright} IUCAA-5/98 \end{flushright}
\n Keywords : Cosmology; L R S Bianchi type-I models; Lyra geometry; Bulk viscous universe\\
\n PACS No. : 98.80
%\newpage
%%%%%%%%%%%%%%%%%%%%%%%%%%%%%%%%%%%%%%%%%%%%%%%%%%%%%%%%%%%%%%%%%%%
\section{Introduction}
%%%%%%%%%%%%%%%%%%%%%%%%%%%%%%%%%%%%%%%%%%%%%%%%%%%%%%%%%%%%%%%%%%%
After Einstein, in 1917, developed his general theory of relativity, in which
gravitation is described in terms of geometry, Weyl, in 1918, proposed a more
general theory in which electromagnetism is also described geometrically. 
However, this theory, based on non-integrability of length transfer, had some 
unsatisfactory features and  did not gain general acceptance. Later Lyra 
\cite{ref1} suggested a modification of Riemannian geometry, which may also
be considered as a modification of Weyl's geometry, by introducing a gauge 
function into the structureless manifold which removes the non-integrability 
condition of the length of a vector under parallel transport and a cosmological 
constant is naturally introduced from the geometry. In subsequent investigations,
Sen \cite{ref2} \& Sen and Dunn \cite{ref3} proposed a new 
scalar-tensor theory of gravitation and constructed an analog 
of the Einstein field equations based on Lyra's geometry.
\par
Halford\cite{ref4} pointed out that the constant displacement vector 
field $\phi_i$ in Lyra's geometry plays the role of a cosmological 
constant in the normal general relativistic treatment. Halford\cite{ref5} 
showed that the scalar-tensor treatment based on Lyra's geometry 
predicts the same effects, within observational limits, as in 
Einstein's theory. Several authors (Bhamra\cite{ref6}, Karade and 
Borikar\cite{ref7}, Kalyanshetti and Wagmode\cite{ref8},
Reddy and Innaiah\cite{ref9}, Beesham\cite{ref10}, Reddy and 
Venkateswarlu\cite{ref11} and Soleng\cite{ref12}) have studied 
cosmological models based on Lyra's geometry with a constant 
displacement field vector. However, this restriction of the 
displacement field to be a constant is a coincidence and there is 
no {\it a priori} reason for it. Singh and Singh{\cite{ref13}}$-${\cite{ref15}} 
and Singh and Desikan\cite{ref16} have studied Bianchi Type I, III, 
Kantowski-Sachs and a new class of models with a time dependent 
displacement field and have made a comparative study of 
Robertson-Walker models with a constant deceleration parameter 
in Einstein's theory with a cosmological term and in the cosmological 
theory based on Lyra's geometry. Recently Pradhan and Vishwakarma\cite{ref17}
investigated a new class of an LRS Bianchi Type-I cosmological models in Lyra 
geometry. Though the displacement vector has no clear and unambiguous interpretations
some efforts have been made to treat the constant displacement vector as the analogue 
of the cosmological constant. Soleng\cite{ref12} has pointed out that 
the cosmologies based on Lyra's manifold with a constant gauge vector 
$\phi$ will either include a creation field and be equal to Hoyle's 
creation field cosmology{\cite{ref18}}$-${\cite{ref20}} or contain a special 
vacuum field which together with the gauge vector may be considered 
as a cosmological term. In the latter case, the solutions are 
same as those of general relativistic cosmologies with a cosmological term.
Recently Behnke {\it et al.}\cite{ref21} also pointed out to an alternative description
of the new cosmological supernova data without a $\Lambda$-term as evidence for
Weyle's geometry of similarity, where Einstein's theory takes the 
form of the conformal-invariant theory of a massless scalar field{\cite{ref22}}
$-${\cite{ref25}. As it has been shown by Weyl already in $1918$, conformal-invariant 
theories correspond to the relative standard of measurement of a conformal-invariant 
ratio of two intervals, given in the geometry of similarity\cite{ref26}
as a manifold of Riemannian geometries connected by conformal transformations.
This ratio depends on nine components of metrics whereas the tenth
component became the scalar dilation field that cannot be removed by the 
choice of the gauge. In the current literature\cite{ref27} this peculiarity
of the conformal-invariant version of Einstein's
dynamics has been overlooked. The energy constraint converts this 
dilation into a time-like classical evolution parameter which scales all 
masses including the Planck mass. In the conformal cosmology (CC),
the evolution of the value of the massless dilation field (in the homogeneous 
approximation) corresponds to that of the scale factor in standard cosmology (SC).
Thus, the CC is a field version of the Hoyle-Narlikar cosmology\cite{ref28}, where 
the redshift reflects the change of the atomic energy levels in the evolution
process of the elementary particle masses determined by the scalar dilation
field\cite{ref23,ref28,ref29}. The CC describes the evolution of the conformal time,
which has a dynamics different from that of the standard Friedmann model.
Behnke {\it et al.}\cite{ref21} have also discussed as an observational argument 
in favour of the CC scenario that the Hubble diagram (effective magnitude-
redshift-relation: $m(z)$) including the recent SCP data\cite{ref26} can
be described without a cosmological constant.  
\par
Most studies in cosmology involve a perfect fluid. However, observed 
physical phenomena such as the large entropy per baryon and the remarkable
degree of isotropy of the cosmic microwave background radiation suggest 
that we should analyse dissipative effects in cosmology. Furthermore, there
are several processes which are expected to give rise to viscous 
effects. These are the decoupling of neutrinos during the radiation 
era and the decoupling of radiation and matter during the recombination
era. Bulk viscosity is associated with the GUT phase transition and 
string creation. The model studied by Murphy\cite{ref38} possessed 
an interesting feature in that the big bang type of singularity of 
infinite space-time curvature does not occur a finite past. However, 
the relationship assumed by Murphy between the viscosity coefficient 
and the matter density is not acceptable at large density. The effect 
of bulk viscosity on the cosmological evolution has been investigated 
by a number of authors in the framework of general theory of 
relativity{\cite{ref30}}$-${\cite{ref45}}. This motivates us to study a 
cosmological viscous fluid model. 
\par
 The Einstein's field equations are a coupled system of highly nonlinear
differential equations and we seek physical solutions to the field equations
for their applications in cosmology and astrophysics. In order to solve the 
field equations we normally assume a form for the matter content or that 
spacetime admits killing vector symmetries\cite{ref46}. Solutions to the field 
equations may also be generated by applying a law of variation for Hubble's 
parameter which was proposed by Berman\cite{ref47}. It is interesting to obverse 
that the law yields a constant value for deceleration parameter (DP). The 
variation of Hubble's law as assumed is not inconsistent with observation 
and has the advantage of providing simple functional forms of the scale 
factor. In simplest case the Hubble law yields a constant value for the DP. 
The cosmological models with constant deceleration parameter (CDP) may be 
divided into two categories. The first category of models with CDP is that 
of models where the cosmic expansion is driven by big bang impulse; all the 
matter and radiation energy is proposed at the big bang epoch and the 
universe has started with singular origin. In the second category of models 
with CDP, the universe has a non-singular origin and the cosmic expansion 
is driven by the creation of matter particles. It is worth observing that 
most of the well-known models of Einstein's theory and Brans-Deke theory 
with curvature parameter $k=0$, including inflationary models, are models 
with CDP. It also measures the deviation from linearity of growth of the 
scale factor. In earlier literature cosmological models with an CDP have 
been studied by Berman\cite{ref47}, Berman and Gomide\cite{ref48}, 
Johri and Desikan\cite{ref49,ref50}, Singh and Desikan\cite{ref16}, 
Maharaj and Naidoo\cite{ref51}, Pradhan {\it et al.}\cite{ref17,ref40,ref42}
and others. This has provided us the motivation to study models with CDP.
\par
In this paper, we have investigated bulk viscous Locally Rotationally 
Symmetric (LRS) Bianchi Type I cosmological models based on Lyra's 
geometry with a time dependent displacement field. It is remarkable
to note here that the time dependent displacement vector may lead to the 
singularity free model\cite{ref52}. We have obtained 
exact solutions of the field equations by assuming the deceleration 
parameter to be constant. The physical behaviour of these models 
have been discussed.
%%%%%%%%%%%%%%%%%%%%%%%%%%%%%%%%%%%%%%%%%%%%%%%%%%%%%%%%%%%%%%%%%%%%%%
%%%%%%%%%%%%%%%%%%%%%%%% SECTION 2 %%%%%%%%%%%%%%%%%%%%%%%%%%%%%%%%%%
\section{Field Equations}
\noindent
The metric for LRS Bianchi Type I spacetime is
\begin{equation}
\label{eq1}
ds^2 = dt^2 - A^2 dx^2 - B^2(dy^2 + dz^2),
\end{equation}
where, $A$ and $B$ are functions of $x$ and $t$. The energy momentum 
tensor in the presence of bulk stress has the form
\begin{equation}
\label{eq2}
T_{ij} = (\rho + \bar{p}) u_i u_j - \bar{p} g_{ij},
\end{equation}
together with comoving coordinates $u^{i} u_{i} = 1$ where $u_{i} = 
(0, 0, 0, 1)$ and 
\begin{equation}
\label{eq3}
\bar{p} = p + \xi u^i_{;i}~{\rm and}.
\end{equation}
Here $\rho,  p, \bar{p}, \xi~{\rm and}~u$ are, respectively, the energy density,
isotropic pressure, effective pressure, bulk viscous coefficient and four-velocity 
vector of the matter distribution. Hereafter, the semi-colon denotes covariant 
differentiation. In general, $\xi$ is a function of time.\\ 
The field equations in normal gauge for Lyra's manifold, as obtained by Sen 
\cite{ref2} are
\begin{equation}
\label{eq4}
R_{ij} - \frac{1}{2} g_{ij} R + \frac{3}{2}\phi_i \phi_j - 
\frac{3}{4} g_{ij}\phi_k \phi^k = - 8\pi G T_{ij},
\end{equation}
where $\phi$ is a time-like displacement field vector defined by 
\begin{equation}
\label{eq5}
\phi_i = (0, 0, 0, \beta(t) ),
\end{equation}
and the other symbols have their usual meaning as in Riemannian geometry. The energy 
momentum tensor $T^{ij}$ is not conserved in Lyra's geometry. Here we want to mention
the fact that the {\it ans\"{a}tz} choosing the coordinate system with matter requir the 
vector field happens to be in the required form exactly in the matter comoving 
coordinates. The essential difference between the cosmological theories based on 
Lyra geometry and the Riemannian geometry lies in the fact that the constant vector 
displacement field $\beta$ arises naturally from the concept of gauge in Lyra geometry 
whereas the cosmological constant $\Lambda$ was introduced in {\it adhoc} fashion in 
the usual treatment.\\ 
The field equations (\ref{eq4}) with the equations (\ref{eq2}) and (\ref{eq5}) 
take the form
\begin{equation}
\label{eq6}
2 \frac{\ddot B}{B} + \frac{\dot B^2}{B^2} - \frac{B^{\prime 2}}{A^2 B^2}
+ \frac{3}{4}\beta^2 = -\chi \bar p,
\end{equation}
\begin{equation}
\label{eq7}
\dot B^{\prime} - \frac{B^{\prime} \dot A}{A} = 0,
\end{equation}
\begin{equation}
\label{eq8}
\frac{\ddot A}{A} + \frac{\ddot B}{B} + \frac{\dot A \dot B}{A B} - 
\frac{B^{\prime\prime}}{A^2 B} + \frac{A^{\prime} B^{\prime}}{A^3 B}
+ \frac{3}{4}\beta^2 = - \chi\bar p,
\end{equation}
\begin{equation}
\label{eq9}
2\frac{B^{\prime\prime}}{A^2 B} - 2 \frac{A^{\prime} B^{\prime}}{A^3 B} 
+ \frac{B^{\prime~2}}{A^2 B^2} - 2\frac{\dot A \dot B}{A B} - 
\frac{\dot B^2}{B^2} + \frac{3}{4}\beta^2 = \chi \rho.
\end{equation}
The energy conservation equation is
\begin{equation}
\label{eq10}
\chi \dot\rho + \frac{3}{2} \beta \dot\beta + \left[ \chi(\rho + \bar p)
+ \frac{3}{2} \beta^2\right]\left(\frac{\dot A}{A} + 2 \frac{\dot B}{B}\right) = 0,
\end{equation}
where $\chi = 8\pi G$. Here and in what follows, a prime and a dot indicate
partial differentiation with respect to $x$ and $t$ respectively. We note 
that the coefficient of bulk viscosity $\xi$ does not appear explicitly 
in the field equations above. For the specification of $\xi$, we assume that
the fluid obeys a barotropic equation of state
\begin{equation}
\label{eq11}
p = \gamma \rho,
\end{equation}
where $\gamma (0\leq\gamma\leq 1)$ is a constant.
%%%%%%%%%%%%%%%%%%%%%%%%%%%%%%%%%%%%%%%%%%%%%%%%%%%%%%%%%%%%%%%%%%%%%%%%%%%%
%%%%%%%%%%%%%%%%%%%%%  SECTION 3  %%%%%%%%%%%%%%%%%%%%%%%%%%%%%%%%%%%%%%%%%%
\section{Solutions of the field equations and discussion}
On integrating the equation (\ref{eq7}), we obtain
\begin{equation}
\label{eq12}
A = \frac{B^{\prime}}{l},
\end{equation}
where $l$ is an arbitrary function of $x$. Using equation (\ref{eq12}),
equations (\ref{eq6}) and (\ref{eq8}) can be written as 
\begin{equation}
\label{eq13}
\frac{B}{B^{\prime}} \frac{d}{dx}\left(\frac{\ddot B}{B}\right) + 
\frac{\dot B}{B^{\prime}}\frac{d}{dt}\left(\frac{B^{\prime}}{B}\right)
+ \frac{l^2}{B^2}\left(1 - \frac{B l^{\prime}}{B^{\prime} l}\right) = 0.
\end{equation}
Since $A$ and $B$ are separable functions of $x$, so, $\frac{B^{\prime}}{B}$
is a function of $x$. Consequently, equation (\ref{eq13}) gives after 
integration
\begin{equation}
\label{eq14}
B = l S(t),
\end{equation}
where $S(t)$ is an arbitrary function of $t$. Using the equation (\ref{eq14}),
(\ref{eq12}) becomes 
\begin{equation}
\label{eq15}
A = \frac{l^{\prime}}{l} S.
\end{equation}
The metric (\ref{eq5}) then takes the form
\begin{equation}
\label{eq16}
ds^2 = dt^2 - S^2 (t)\left[ dX^2 + e^{2X}(dy^2 + dz^2)\right],
\end{equation}
where $X = {\rm ln}~ l$. Equations (\ref{eq6}) and (\ref{eq9}) give
\begin{equation}
\label{eq17}
\chi\bar p = \frac{1}{S^2} - 2\frac{\ddot S}{S} - \frac{\dot S^2}{S^2} - \frac{3}{4}\beta^2,
\end{equation}
\begin{equation}
\label{eq18}
\chi\rho = 3 \frac{\dot S^2}{S^2} - \frac{3}{S^2} - \frac{3}{4}\beta^2.
\end{equation}
\noindent
Using the equation (\ref{eq11}) and eliminating $\rho(t)$ from equations
(\ref{eq17}) and (\ref{eq18}), we have
\begin{equation}
\label{eq19}
2\frac{\ddot S}{S} + (1 + 3\gamma) \frac{\dot S^2}{S^2} - 
(1 + 3\gamma)\frac{1}{S^2} + \frac{3}{4} (1 - \gamma) \beta^2 = 3 \chi \xi \frac{\dot S}{S}.
\end{equation}
In most of the investigations involving bulk viscosity, the coefficient of
bulk viscosity is assumed to be a simple power function of the energy density
{\cite{ref35}}$-${\cite{ref37}} 
\begin{equation}
\label{eq20}
\xi(t) = \xi_0 \rho^n,
\end{equation}
where $\xi_0$ and $n$ are constants. If $n=1$, equation (\ref{eq20}) may correspond to a 
radiative fluid. However, more realistic models are based on $n$ lying in the regime 
$0\leq n \leq \frac{1}{2}$. 
\par
Using equation (\ref{eq20}), (\ref{eq19}) can be written as
\begin{equation}
\label{eq21}
2 \frac{\ddot S}{S} + (1 + 3\gamma) \frac{\dot S^2}{S} - (1 + 3\gamma) \frac{1}{S^2}
 + \frac{3}{4}(1 - \gamma)\beta^2 = 3 \chi \xi_0 \rho^n \frac{\dot S}{S}.
\end{equation}
In order to obtain the explicit dependence of $S(t)$ on time, we assume 
the deceleration parameter to be constant, i.e., 
\begin{equation}
\label{eq22}
q = - \frac{S \ddot S}{\dot S^2} = - \left(\frac{\dot H + H^2}{H^2}\right) 
= b~ (\rm constant),
\end{equation}
where $H = \frac{\dot S}{S}$ is the Hubble parameter. It is worthwhile to note 
that the solutions arrived at and the problem scanned cannot be properly
handled by taking the DP to be a variable one. By considering DP to be a 
function of time $t$, we observe that $q$ is rapidly increasing which is
inconsistent and hence we should restrict ourselves to CDP. The equation 
(\ref{eq22}) can be rewritten as 
\begin{equation}
\label{eq23}
\frac{\ddot S}{S} + b \frac{\dot S^2}{S^2} = 0.
\end{equation}
Integration of equation (\ref{eq23}) gives us the exact solution
\begin{eqnarray}
\label{eq24}
S(t) & = & \left[a(t - t_0)\right]^{\frac{1}{1+b}},~~~{\rm when}~b\neq -1\\
{~~} & = & m_1 e^{m_2 t}\nonumber,~~~{\rm when}~~~b = -1\nonumber
\end{eqnarray}
where $a, m_1$ and $m_2$ are constants of integration and the constant
$t_0$ allows us the freedom of choosing the initial time. 
\newline
On using equation (\ref{eq23}) in (\ref{eq21}), we obtain
\begin{equation}
\label{eq25}
\left[3(1 + \gamma) - 2(1 + b)\right] H^2 - (1 + 3\gamma) \frac{1}{S^2}
+ \frac{3}{4} (1 - \gamma) \beta^2 = 3 \chi \xi_0 \rho^n H.
\end{equation}
Equation (\ref{eq25}) with equation (\ref{eq18}) leads to
\begin{eqnarray}
\label{eq26}
\left[ 3(1 + \gamma) - 2(1 + b)\right] H^2 & - & (1 + 3\gamma) \frac{1}{S^2} +\frac{3}{4} (1 - \gamma) \beta^2\nonumber\\ 
& = & 3 \chi^{1 - n}\xi_0 H \left[3 H^2 - \frac{3}{S^2} - \frac{3}{4}\beta^2\right]^n .
\end{eqnarray}
\newline
\newline
In what follows, we will solve the equations for two particular values of
$n$, viz., $n = 0$ and $n = 1$.
%%%%%%%%%%%%%%%%%%%%%%%%%%%%%%%%%%%%%%%%%%%%%%%%%%%%%%%%%%%%%%%%%%%%%%%%%
%%%%%%%%%%%%%%%%%%%%%%%   SUBSECTION 3.1  %%%%%%%%%%%%%%%%%%%%%%%%%%%%%%%
\subsection{Solutions for $\xi = \xi_0$}
In this case, we assume $n = 0$ in equation (\ref{eq20}). Equations 
(\ref{eq20}) and (\ref{eq26}) become $\xi = \xi_0 = {\rm constant}$ and
\begin{equation}
\label{eq27}
\left[ 3(1 + \gamma) - 2(1 + b)\right] H^2 - (1 + 3\gamma)\frac{1}{S^2}
 + \frac{3}{4} (1 - \gamma) \beta^2 = 3\chi\xi_0 H . 
\end{equation}

\noindent
{\bf Case (i)}:~$b\neq -1$. For singular models, $S(0) = 0$ and hence,
equation (\ref{eq24}) leads to 
\begin{equation}
\label{eq28}
S = a~ t^{\frac{1}{(1 + b)}} . 
\end{equation}
In the figure 1, we have shown how the scale factor evolves with
time for two choices of $b$, $b = 1$ and $b = 2$.
\begin{figure}[ht] %ORIGINAL SIZE: width=1.4TRUEIN; height=1.5TRUEIN
\vspace*{13pt}
\centerline{\includegraphics[width=1.0\textwidth,angle=0]{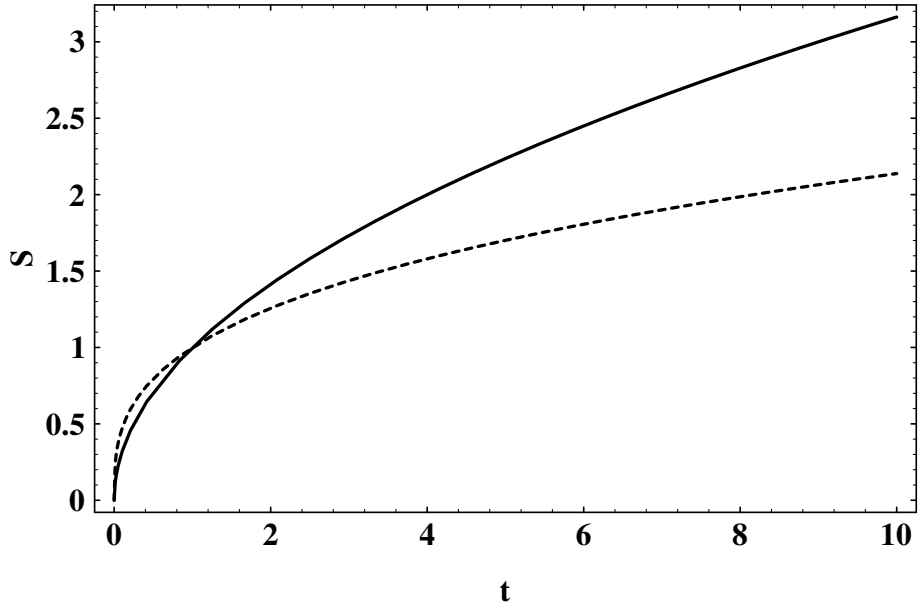}} %100 percent
\vspace*{13pt}
\caption{The behaviour of scale factor for MODEL I ($b\neq -1$) with time 
for $b = 1$ (solid line) and $b = 2$ (dashed line) for $a = 1$.}
\end{figure}
\newline
From equation (\ref{eq28}), we have
\begin{equation}
\label{eq29}
H = \frac{1}{(1+b)t}.
\end{equation}
The evolution of the Hubble parameter with time is shown in figure 2 
for $b = 1$ and $b=2$ for $a = 1$.
\begin{figure}[htb] %ORIGINAL SIZE: width=1.4TRUEIN; height=1.5TRUEIN
\vspace*{13pt}
\centerline{\includegraphics[width=1.0\textwidth,angle=0]{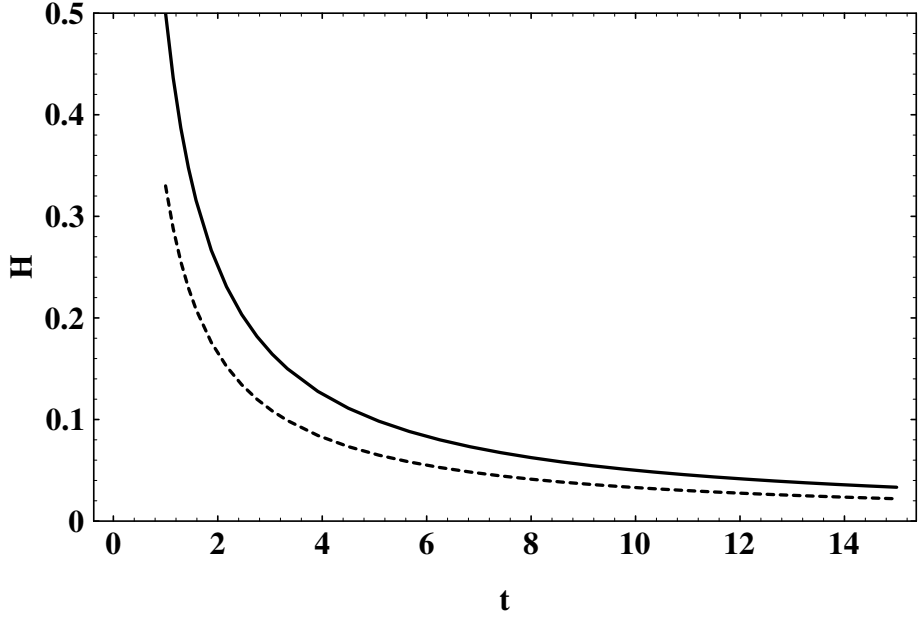}}
%%\centerline{\psfig{file=fig2.eps,width=8cm}} %100 percent
\vspace*{13pt}
\caption{The behaviour of Hubble parameter for MODEL I ($b\neq -1)$ with time 
for $b = 1$ (solid line) and $b = 2$ (dashed line).}
\end{figure}
\newline
\noindent
Using the two equations (\ref{eq28}) and (\ref{eq29}) in the 
equation (\ref{eq27}), we obtain
\begin{eqnarray}
\label{eq30}
\beta^2 = \frac{4}{3}\frac{1}{(1 - \gamma)(1 + b)^2 t^2}\times \qquad\qquad\qquad\qquad\qquad\qquad\qquad\qquad \nonumber\\
\left[ 3 \chi \xi_0 (1 + b)t - (3\gamma - 2 b + 1) + \frac{(3\gamma + 1)(1 + b)^2}{a^2 t^{-\frac{2b}{(1+b)}}}\right] .
\end{eqnarray}
Using equation (\ref{eq30}) in (\ref{eq18}), we get
\begin{equation}
\label{eq31}
\chi \rho = \frac{1}{(1 - \gamma)(1 + b)^2 t^2}\left[2(2-b) - 3\chi\xi_0(1+b)t\right]
 - \frac{4}{(1 - \gamma) a^2 t^{\frac{2}{(1+b)}}}.
\end{equation}
\begin{figure}[ht]
\centerline{\includegraphics[width=1.0\textwidth,angle=0]{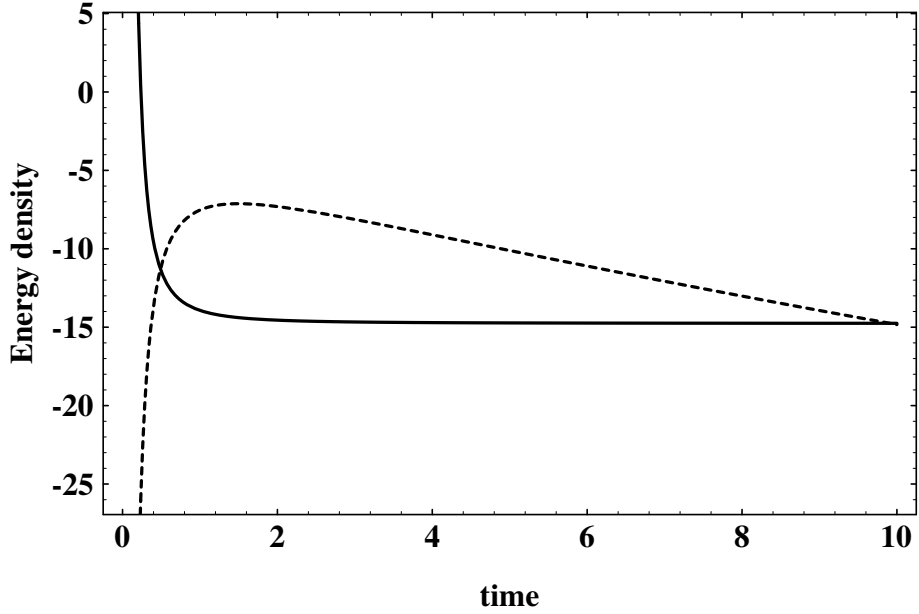}}
%%\centerline{\psfig{file=fig3.eps,width=8cm}} %100 percent
\vspace*{13pt}
\caption{The plot of energy density for MODEL I ($b\neq -1)$ with time for $b = 1$ (solid line) 
and $b = 2$ (dashed line). (Here, we have taken $\chi = 1$, $\gamma = 0.4$ and $\xi_0 = 0.5$)}
\end{figure}
\newline
The geometry of the universe, in this case, is described by the
line-element 
\begin{equation}
\label{eq32}
ds^2 = dt^2 - a^2 t^{\frac{2}{(1+b)}} \left[dX^2 + e^{2X}(dy^2 + dz^2)\right].
\end{equation}
In equation (\ref{eq30}), we observe that $\beta^2 = 0$ when $t = t_c$, 
where the critical time $t_c$ satisfies 
\begin{eqnarray}
t_c^{-\frac{2b}{(1+b)}}\left[(3\gamma - 2b + 1) + 3\gamma \xi_0(1+b) t_c\right] 
= \frac{(3\gamma + 1)(1 + b)^2}{a^2}.\nonumber
\end{eqnarray}
It is also seen that $\beta^2 > 0$ provided $t > t_c$ and $\beta^2 < 0$ 
provided $t < t_c$. 
\par
From equation (\ref{eq31}), it is also observed that $\rho > 0$ provided
$t > T_c$, where $T_c$ is the critical time given by
\begin{eqnarray}
T_c^{-\frac{2b}{(1+b)}}\left[2(2-b) - 3\chi \xi_0 (1+b) T_c\right] 
= \frac{4 (1+b)^2}{a^2}.\nonumber
\end{eqnarray}
The figure 3 shows the evolution of the energy density with time for
two values of $b$, viz., $b = 1$ and $b = 2$. We see from the figure that
when $t > T_c$, the energy density is positive, but when $t < T_c$, the 
energy density becomes negative and then evolves to a constant negative value.
\newline
\newline
{\bf Physical behaviour of the model:}\\
In the case of a non-flat model, when $b\neq -1$, the Ricci scalar becomes
\begin{equation}
\label{eq33}
R = \frac{1}{a^2 t^{\frac{2}{(1+b)}}} - \frac{(1-b)t}{(1+b)}.
\end{equation}

In equation (\ref{eq33}), we see that when $t\rightarrow 0$; 
(i) $R\rightarrow\infty$ if $b = 0$, and (ii) $R\rightarrow\infty$ if
$b\geq 1$. The equation (\ref{eq33}) also suggests that when 
$t\rightarrow\infty$, $R\rightarrow 0$ if $b\geq 0$. The expansion and 
shear scalars are given by
\begin{equation}
\label{eq34}
\theta = \frac{3}{(1+b)t},~~\sigma = 0,
\end{equation}
respectively.
\par
The model has a singularity at $t = 0$. As $t\rightarrow \infty$, the 
expansion ceases. In this model, $\frac{\sigma}{\theta} = 0$, which
confirms the isotropic nature of the spacetime which we have obtained
in (\ref{eq32}). 

\noindent
{\bf Case (ii)}:~$b=-1$. In this case, we obtain from the equation (\ref{eq22})
\begin{equation}
\label{eq35}
\dot H = 0 \Rightarrow H = H_0 = {\rm constant,}
\end{equation}
and scale factor is given by 
\begin{eqnarray}
S = m_1 e^{m_2 t}.\nonumber
\end{eqnarray}
The evolution of the scale factor with time in this case for $m_1 = 1$ and
$m_2 = 2, 3$ is shown in figure 4.
\begin{figure}[ht]
\centerline{\includegraphics[width=1.0\textwidth,angle=0]{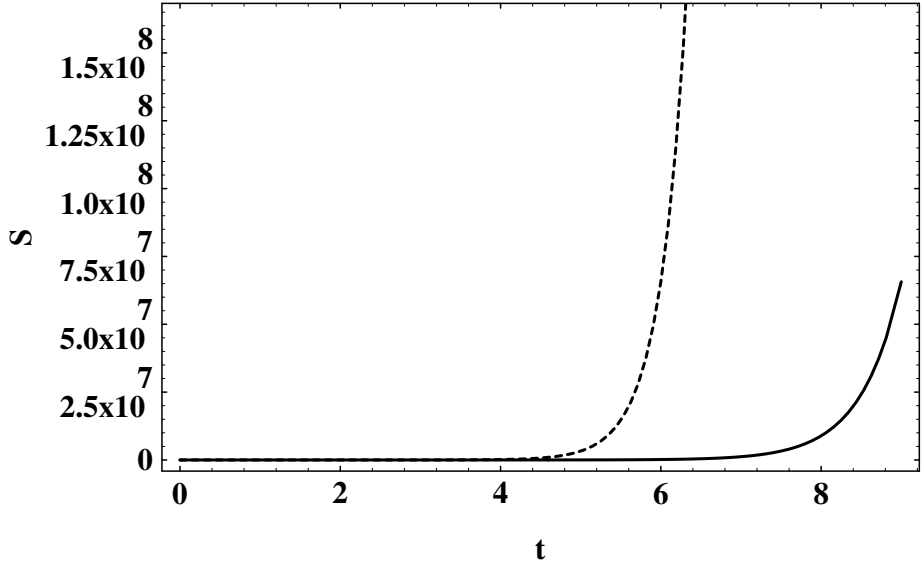}}
%%\centerline{\psfig{file=fig4.eps,width=8cm}} %100 percent
\vspace*{13pt}
\caption{The behaviour of the scale factor for MODEL I ($b = -1)$ with time for 
$m_2 = 2$ (solid line) and $m_2 = 3$ (dashed line).}
\end{figure}
\newline
On using equation (\ref{eq35}), equations (\ref{eq27}) and (\ref{eq18}) 
reduce to 
\begin{equation}
\label{eq36}
\beta^2 = \frac{4}{3 (1 - \gamma)}\left[ 3 \chi \xi_0 H_0 - 3 (1 + \gamma) H_0^2 + 
\frac{(1 + 3\gamma)}{m_1^2}e^{-2 m_2 t}\right],
\end{equation}
\begin{equation}
\label{eq37}
\chi \rho = \frac{1}{(1 - \gamma)}\left[ 6 H_0^2 - 3 \chi \xi_0 H_0 
- \frac{4}{m_1^2} e^{-2 m_2 t}\right].
\end{equation}
\begin{figure}[h] %ORIGINAL SIZE: width=1.4TRUEIN; height=1.5TRUEIN
\vspace*{13pt}
\centerline{\includegraphics[width=1.0\textwidth,angle=0]{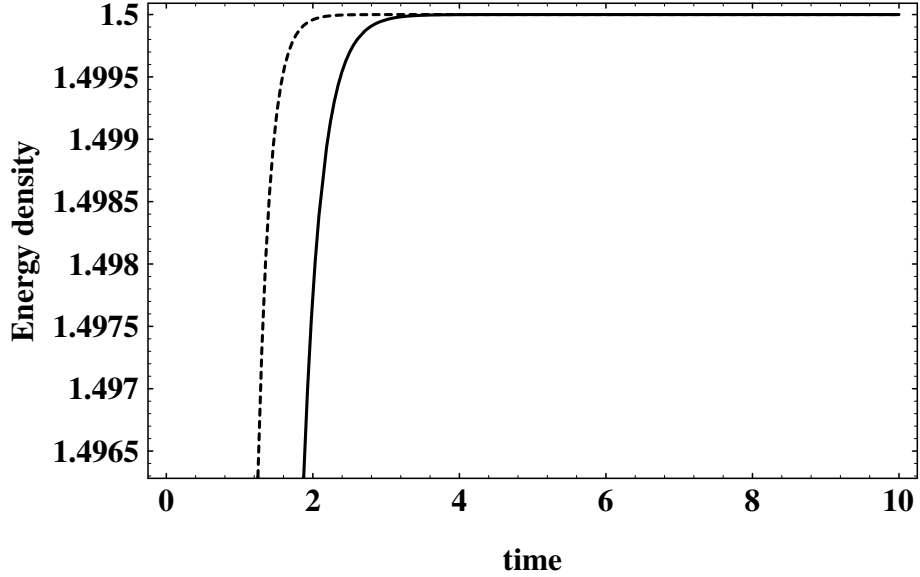}}
%%\centerline{\psfig{file=fig5.eps,width=8cm}} %100 percent
\vspace*{13pt}
\caption{The behaviour of the energy density for MODEL I ($b = -1)$ with time 
for $m_2 = 2$ (solid line) and $m_2 = 3$ (dashed line). (Here, we have taken 
$\chi = 1$, $m_1 = 1$, $\gamma = 0.4$, $\xi_0 = 0.4$ and $H_0 = 0.5$)}
\end{figure}
The geometry of the Universe, in this case, is described by 
the line element
\begin{equation}
\label{eq38} 
ds^2 = dt^2 - a^2 e^{2 H_0 t} \left[ dX^2 + e^{2X} (dy^2 + dz^2)\right].
\end{equation}
\noindent
From equation (\ref{eq36}), we see that 
\begin{enumerate}
\item $\beta^2 > 0$ provided $\xi_0 < \frac{2 H_0}{\chi}$ for all $t > T$,
\item $\beta^2 > 0$ provided $\xi_0 > \frac{2 H_0}{\chi}$ for all $t < T$,
\item $\beta^2 < 0$ provided $\xi_0 > \frac{2 H_0}{\chi}$ for all $t > T$,
\item $\beta^2 < 0$ provided $\xi_0 < \frac{2 H_0}{\chi}$ for all $t < T$,
\item $\beta^2 = 0$ provided $t = T$,
\end{enumerate}
where
\begin{eqnarray}
T= \frac{1}{2 m_2}{\rm ln}\left[\frac{(1 + 3\gamma)}{m_1^2\{3 (1 + \gamma) H_0^2 
- 3\chi\xi_0 H_0\}}\right].\nonumber
\end{eqnarray}
From equation (\ref{eq37}), we observe that
\begin{eqnarray}
(i)&\rho > 0,& {\rm if}~\xi_0 < \frac{2 H_0}{\chi} ~{\rm for~ all}~t > T_1,\nonumber\\
(ii)&\rho > 0,& {\rm if}~\xi_0 > \frac{2 H_0}{\chi} ~{\rm for~ all}~0 < t < T_1,\nonumber
\end{eqnarray}
where, 
\begin{eqnarray}
T_1 = \frac{1}{2 m_2} ~{\rm ln}\left[ \frac{4}{m_1^2\{6 H_0^2 - 3\chi \xi_0 H_0\}}\right].\nonumber
\end{eqnarray}
In the figure 5, we have shown the evolution of the energy density 
with time for two values of $m_2$, viz., $m_2 = 2$ and $m_2 = 3$. We see 
from the figure that the energy density increases with time and then 
approaches a constant positive value.
\newline
{\bf Physical behaviour of the model:}\\
The Ricci scalar $R$ is given by
\begin{equation}
\label{eq39}
R = 2 H_0^2 - \frac{1}{a^2 e^{2 H_0 t}}.
\end{equation}
\noindent
We clearly observe from equation (\ref{eq39}) that (i) when $t\rightarrow 0, 
R\rightarrow\left(2 H_0^2 - \frac{1}{a^2}\right)$, and (ii) when 
$t \rightarrow\infty, R\rightarrow 2 H_0^2$. The expansion and shear 
scalars are
\begin{equation}
\label{eq40}
\theta = 3 H_0,~ \sigma = 0.
\end{equation}
This model represents an uniform expansion as can be seen from 
equation (\ref{eq40}). The flow of the fluid is geodetic as the 
acceleration vector $f_i = (0, 0, 0, 0)$. 
%%%%%%%%%%%%%%%%%%%%%%%%%%%%%%%%%%%%%%%%%%%%%%%%%%%%%%%%%%%%%%%%%%%%
%%%%%%%%%%%%%%%%%%%%%%%%%%%  SUBSECTION 3.2  %%%%%%%%%%%%%%%%%%%%%%%
\subsection{Solutions for $\xi = \xi_0 \rho$}
In this case, we choose $n = 1$ and hence, equations (\ref{eq20}) and (\ref{eq26})
now reduce to 
\begin{equation}
\label{eq41}
\xi = \xi_0 \rho,
\end{equation}
\[
\left[ 3(1 + \gamma) - 2 (1 + b)\right] H^2 - (1 + 3\gamma) \frac{1}{S^2} + 
\frac{3}{4} (1 - \gamma) \beta^2 = \]
\begin{equation}
\label{eq42}
3 \xi_0 H \left(3 H^2 - \frac{3}{S^2} - 
\frac{3}{4}\beta^2\right).
\end{equation}
respectively.
\newline
Equation (\ref{eq42}) can be written as
\begin{equation}
\label{eq43}
\beta^2 = \frac{4}{3} ~~\frac{9 \xi_0 H^3 - (3 \gamma - 2 b + 1) H^2 + 
(1 + 3\gamma - 9\xi_0 H) S^{-2}} {(1 - \gamma + 3 \xi_0 H)}.
\end{equation}
\noindent
{\bf Case (i)}:~$b\neq -1$.\\
Here we are considering a singular model. Hence, using equations (\ref{eq28}) and
(\ref{eq29}) in equations (\ref{eq43}) and (\ref{eq18}), we obtain
\[
\beta^2 = \frac{4}{3 a^{2} (1 + b)^{2} t^{2}[(1 - \gamma)(1 + b)t + 3\xi_0] }\times
\]
\begin{equation}
\label{eq44}
[9 a^2 \xi_0 - 9\xi_0 (1 + b)^2 t^{\frac{2 b}{(1+b)}} - (3 \gamma - 2 b + 1)(1 + b)
 a^2 t + (1 + 3\gamma) (1 + b)^3 t^{\frac{3b+1}{1+b}}]
\end{equation}
\begin{equation}
\label{eq45}
\chi\rho = \frac{2}{a^2 (1 + b) t}~~\frac{[(2 - b) a^2 - 2 (1 + b)^2 t^{\frac{2 b}
{(1+b)}}]} {[ (1 - \gamma) (1 + b) t + 3 \xi_0]}.
\end{equation}
\begin{figure}[h] %ORIGINAL SIZE: width=1.4TRUEIN; height=1.5TRUEIN
\vspace*{13pt}
\centerline{\includegraphics[width=1.0\textwidth,angle=0]{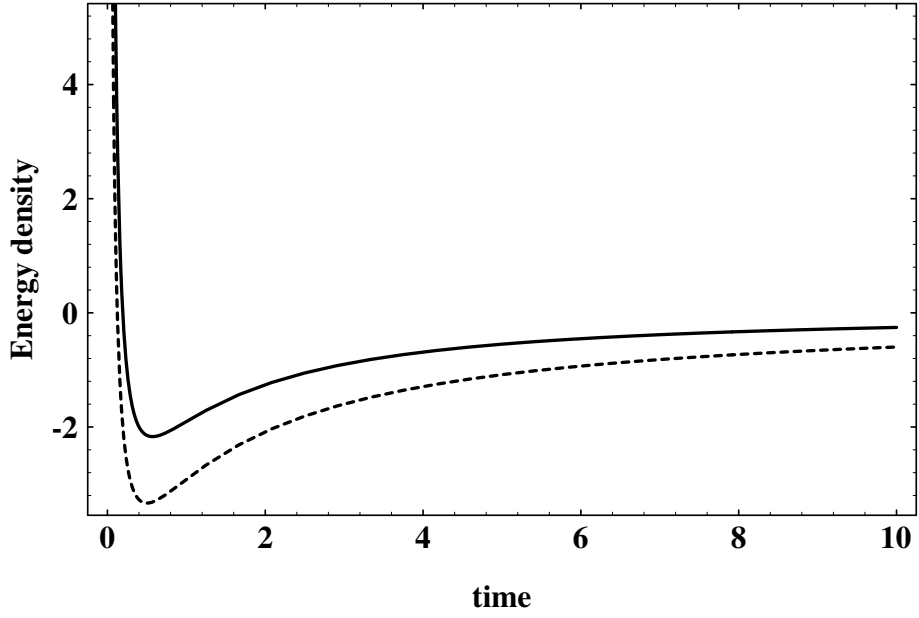}}
%%\centerline{\psfig{file=fig6.eps,width=8cm}} %100 percent
\vspace*{13pt}
\caption{The behaviour of the energy density for MODEL II ($b\neq -1)$ with time 
for $b = 0.5$ (solid line) and $b = 1$(dashed line). (Here, we have taken
$\chi = 1$, $a = 1$, $\gamma = 0.4$ and $\xi_0 = 0.4$)}
\end{figure}
Therefore, in this case, from equation (\ref{eq44}), we see that
\begin{enumerate}
\item $\beta^2 = 0$ if $t = t_c$,
\item $\beta^2 > 0$ if $t = t_c$,
\item $\beta^2 < 0$ if $t < t_c$,
\end{enumerate}
where, the critical time $t_c$ satisfies the relation
\begin{eqnarray}
(1 + b)^2~t_c^{\frac{2 b}{(1+b)}} [9 \xi_0 - (1+3\gamma)(1+b) t_c] = 9 a^2
 \xi_0 - (3\gamma - 2 b + 1)(1 + b)a^2 t_c\nonumber.
\end{eqnarray}
It can also be seen from equation (\ref{eq45}) that $\rho > 0$ if $b<2$ 
with $t > 0$. The behaviour of the energy density with time is shown
in the figure 6 for $b = 0.5$ and $b = 1$. We see from the figure 
that the energy density approaches a small positive value after 
initially becoming negative for a short period of time. The physical 
properties of this model are similar to those discussed for model I.\\ 
\noindent
{\bf Case (ii)}:~$b=-1$. \\
As in the previous model, we now have 
\begin{equation}
\label{eq46}
H = H_0 = {\rm constant}.
\end{equation}
Equations (\ref{eq43}) and (\ref{eq18}) become 
\begin{equation}
\label{eq47}
\beta^2 = \frac{4}{3}~~\frac{[9 \xi_0 m_1^2 H_0^3 - 3 (1 + \gamma)H_0^2 m_1^2 + 
(1 + 3 \gamma - 9\xi_0 H_0)e^{-2 m_2 t}]}{[1 - \gamma + 3 \xi_0 H_0] m_1^2},
\end{equation}
\begin{equation}
\label{eq48}
\chi\rho = \frac{2}{(1 - \gamma + 3\xi_0 H_0) m_1^2} \left[ (3 \gamma - 1)
e^{-2 m_2 t} - 3 \gamma m_1^2 H_0^2\right].
\end{equation}
\begin{figure}[h] %ORIGINAL SIZE: width=1.4TRUEIN; height=1.5TRUEIN
\vspace*{13pt}
\centerline{\includegraphics[width=1.0\textwidth,angle=0]{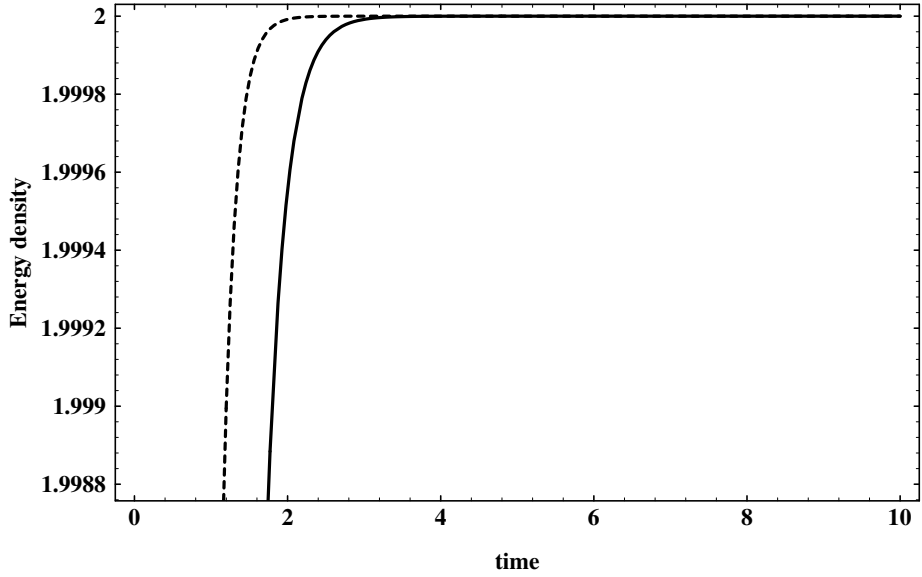}}
%%\centerline{\psfig{file=fig7.eps,width=8cm}} %100 percent
\vspace*{13pt}
\caption{The behaviour of the energy density for MODEL II ($b = -1)$ with time 
for $m_2 = 2$ (solid line) and $m_2 = 3$ (dashed line). (Here, we have taken
$m_1 =1$, $\gamma = 0.4$, $\xi_0 = -0.6$, and $H_0 = 0.5$)}
\end{figure}
In equation (\ref{eq47}), we observe that
\begin{enumerate}
\item $\beta^2 > 0$ if $\xi_0 < \frac{(1+\gamma)}{3 H_0}$ for all $t < T_2$,
\item $\beta^2 > 0$ if $\xi_0 > \frac{(1+\gamma)}{3 H_0}$ for all $t > T_2$,
\item $\beta^2 < 0$ if $\xi_0 < \frac{(1+\gamma)}{3 H_0}$ for all $t > T_2$,
\item $\beta^2 < 0$ if $\xi_0 > \frac{(1+\gamma)}{3 H_0}$ for all $t < T_2$,
\item $\beta^2 = 0$ if $\xi_0 \neq \frac{(1+\gamma)}{3 H_0}$ for all $t = T_2$,
\end{enumerate}
where 
\begin{eqnarray}
T_2 = \frac{1}{2 m_2} {\rm ln}\left[\frac{(1 + 3 \gamma - 9\xi_0 H_0)}
{3 m_1^2 H_0^2 (1 + \gamma - 3 \xi_0 H_0)}\right]\nonumber.
\end{eqnarray}
From the equation (\ref{eq48}), we see that
\begin{eqnarray}
(i)&\rho > 0,& {\rm provided}~\xi > \frac{(\gamma - 1)}{3 H_0} ~{\rm for ~all}~
t > T_3,\nonumber\\
(ii)&\rho > 0,& {\rm provided}~\xi < \frac{(\gamma - 1)}{3 H_0} ~{\rm for ~all}
~t < T_3,\nonumber
\end{eqnarray}
where 
\begin{eqnarray}
T_3 = \frac{1}{2 m_2}{\rm ln}\left[\frac{3\gamma - 1}{3 \gamma m_1^2 H_0^2}\right]\nonumber.
\end{eqnarray}
The behaviour of the energy density with time is shown in figure 7.
We see that the energy density increases with time and approaches a 
constant positive value. The physical properties of this model 
is similar to those of model I. 

%%%%%%%%%%%%%%%%%%%%%%%%%%%%%%%%%%%%%%%%%%%%%%%%%%%%%%%%%%%%%%%%%%%%%%%%%%%
      
%%%%%%%%%%%%%%%%%%%%%  SECTION 4 %%%%%%%%%%%%%%%%%%%%%%%%%%%%%%%%%%%%%%%
\section{Discussion and Conclusion}
In this paper, we have investigated LRS Bianchi type I models with a 
bulk viscous fluid and obtained exact solutions for a constant 
deceleration parameter. We have assumed the coefficient of bulk
viscosity to be of the form $\xi(t) = \xi_0 \rho^n$; where $\rho$ is 
the energy density and $n$ is the power index. The behaviour of the
displacement field $\beta$ and the energy density have been examined 
for values of $n = 0 $ and $n = 1$ for both (i) power-law and (ii) 
exponential expansion of a non-flat universe. The model discussed here is
isotropic and homogeneous and in view of the assumption of isotropy, the
shear viscosity cannot exist. The effect of bulk viscosity is to 
introduce a change in the perfect fluid model.\\
Recently there is an upsurge of interest in scalar fields in general relativity and 
alternative theories of gravitation in the context of inflationary 
cosmology{\cite{ref53}}$-${\cite{ref55}}. Therefore the study of cosmological
models in Lyra geometry may be relevant for inflationary models.
Further the space dependence of the displacement field $\beta$ is important for 
inhomogeneous models for the early stage of the evolution of the universe. Besides,
the implication of Lyra's geometry for astrophysical interesting bodies is still an open 
question. The problem of equations of motion and gravitational radiation need
investigation. Finally in spite of very good possibility for Lyra's geometry to provide 
a theoretical foundation for Relativistic Gravitation, Astrophysics and Cosmology,
the experimental point is yet to be undertaken. But still the theory needs a fair trial.     
\section*{Acknowledgement}
One of the authors (AP) wishes to thank the Harish Chandra Research Institute, 
Allahabad, India for warm hospitality and excellent facilities where this work 
was done. We would like to thank Kalyani Desikan for many helpful discussions
on the first draft of this paper.

\end{document}